\documentclass[12pt]{article}
\usepackage{hyperref,amsfonts,amssymb,theorem}

\newtheorem{statement}{Statement}

\title{On three-dimensional quasi-St\"ackel Hamiltonians}
\author{V.G. Marikhin \footnote{e-mail: mvg@itp.ac.ru}\\
L.D. Landau Institute for Theoretical Physics, RAS\\
Chernogolovka, Moscow region, Russia}
\begin{document}
\maketitle
\thispagestyle{empty}
\begin{abstract}
A three-dimensional integrable generalization of the St\"ackel
systems is proposed. The classification of such systems is obtained which
results in two families. The first one is the direct sum of the two-dimensional
case which is equivalent to the representation of the Schottky--Manakov top in
the quasi-St\"ackel form and a St\"ackel one-dimensional system. The second
family is probably a new three-dimensional system. The system of hydrodynamic
type, which we get from this family in a usual way, is a 3-dimensional
generalization of the Gibbons--Tsarev system. A generalization of the
quasi-St\"ackel systems to the case of any dimension is discussed.
\end{abstract}



\vspace{7mm}

\section{St\"ackel Hamiltonians}
\setcounter{equation}{0}
The St\"ackel systems \cite{stack} were considered in many works, see e.g. \cite{eiz}, \cite{blaz} and references therein, \cite{blaz2}.
Associated with this approach questions about the related problems on the separation of variables were considered for example in the works \cite{skl},\cite{kaln}.
The popularity of these systems can be explained by the fact that a number of important finite-dimensional integrable systems can be brought to the St\"ackel form by a change of variables.
The goal of this article is the study of quasi-St\"ackel systems which can be described as integrable
deformations of St\"ackel systems with the same principal parts.

First, we recall the general construction of the classical St\"ackel system with $n$ degrees of freedom. The corresponding Hamiltonians are defined by the system of linear algebraic equations.
\begin{equation}\label{stack}
\sum\limits_{k=0}^{n-1} S^{-1}_{ik}H_k=\psi (p_i,q_i)\Rightarrow H_{k}=\sum\limits_{i=1}^n S_{ki}\psi (p_i,q_i),\quad S^{-1}_{ik}=f_k(p_i,q_i).
\end{equation}
The matrix $S_{ki}$ is called the St\"ackel matrix. Let us consider the simplest inverse St\"ackel matrix $S_{ik}^{-1}=q_{i}^{n-k-1},\quad i=1,\dots ,n,\;k = 0,\dots ,n-1$ (this is the so-called Benenti case \cite{benenti},\cite{benenti2}) and the function  $\psi (p_k,q_k)=s(q_k)p_k^2+u(q_k).$ This approach is usual to obtain representation of the St\"ackel system in terms of separated variables.
Solving of the system (\ref{stack}) yields a family of Hamiltonians which is commutative with respect to the canonical Darboux--Poisson bracket $\{q_i,p_j\}=\delta_{ij}$:
\begin{equation}
H(\alpha)=\sum\limits_{k=1}^n (s(q_k)p_k^2+u(q_k))\prod\limits_{j\neq k}\frac{\alpha-q_j}{q_k-q_j},\quad \lbrace H(\alpha),H(\beta)\rbrace=0.
\end{equation}
The coefficients of the generating function $H(\alpha)=\sum\limits_{k=0}^{n-1}H_k\alpha^{n-k-1}$ provide a set of $n$ Hamiltonians in involution.
Therefore the St\"ackel system described by these Hamiltonians is integrable in the Liouville sense.

\section{Quasi-St\"ackel Hamiltonians in $n=3$ case}
The quasi-St\"ackel systems in the two-dimensional case were introduced in \cite{umn}. The analogous systems were considered in \cite{Gram}, \cite{ferap}, \cite{Ye}, see also references therein.
We recall that the Hamiltonians of quasi-St\"ackel systems are Hamiltonians of Stackel systems plus magnetic (linear in momenta) terms.

The Hamiltonians $h_i$ of quasi-St\"ackel systems in the Benenti case are defined for arbitrary $n$ as follows:
\begin{equation}\label{q-stack}
\sum\limits_{k=0}^{n-1}q_i^k h_{n-1-k}=S(q_i)p_i^2+\sum\limits_{j=1}^nz_{i,j}(\vec{q})p_j+u_i(\vec{q}).
\end{equation}
One can obtain a generating function $h(\alpha)$ analogously to the St\"ackel case
\begin{equation}\label{ha}
h(\alpha)=\sum\limits_{k=1}^n \Bigl( S(q_k)p_k^2+\sum\limits_{i=1}^nz_{k,i}(\vec{q})p_i+u_k(\vec{q})\Bigr) \prod\limits_{j\neq k}\frac{\alpha-q_j}{q_k-q_j}.
\end{equation}

In order to characterize the integrable cases, we find the Hamiltonians $h_i$ as the coefficients of the generating function $h(\alpha)=\sum\limits_{k=0}^{n-1}h_k\alpha^{n-k-1}$ (\ref{ha}), compute the commutators and equate them to zero:
\begin{equation}\label{hh}
\lbrace h_i,h_j \rbrace=0,\; i,j=0,\dots ,n-1.
\end{equation}

Unlike the St\"ackel case these commutators do not vanish identically and in order to obtain a full classification one has
to determine the functions $S,z_{i,j},u_i$ from this set of $\frac{n(n-1)}{2}$ equations which are quadratic in the momenta.
This is a difficult problem for $n>3$ and we consider only the case $n=3$ in this article.

The first step is to solve the equations corresponding to vanishing of the quadratic in the momenta terms.
This can be done by straightforward, although tedious computation and we arrive to the following statement.

\begin{statement}
If the family (\ref{ha}) is commutative then the coefficients $z_{ij}$
can be brought to the form
\begin{equation}\label{q-stack2}
z_{ij}=\Delta_{ij}\frac{\sqrt{S(q_i)}\sqrt{S(q_j)}}{q_i-q_j},\quad z_{ii}=0
\end{equation}
where $\Delta_{ij}$ are some constants.
\end{statement}

Notice that, strictly speaking, the diagonal coefficients may be of the form $z_{ii}=S(q_i)\frac{\partial}{\partial_{q_i}}F(\vec{q})$, but one can set  $z_{ii}=0$ by use of the canonical transformation $p_i\rightarrow p_i-\frac{1}{2}\frac{\partial}{\partial_{q_i}}F(\vec{q})$.

The analysis of the rest equations brings to the following two (non-St\"ackel) cases.

1. Symmetric case: $\Delta_{i,i}=0,\;\Delta_{i\neq j}=-\delta,\;S(x)=a(x)^2$ where $a(x)$ is a quadratic polynomial.

2. Non-symmetric case: $\Delta_{2,3}=\Delta_{3,2}=-\delta$ and all other $\Delta_{i,j}=0$ (up to change of numeration) and $S$ is a polynomial of sixth degree.

These cases describe all possible three-dimensional quasi-St\"ackel systems.

We recall that the dynamics of the quasi-St\"ackel system is defined by Hamiltonians $h_i.$
Now we consider these two cases in detail.

\subsection{Symmetric case}
In the general symmetric case the final answer is given by the following statement.

\begin{statement}
In the symmetric case 1 the
resulting commutative family of Hamiltonians is of the form (\ref{ha})
\begin{equation}\label{ha2}
h(\alpha)=\sum\limits_{k=1}^3 \Bigl( S(q_k)p_k^2+\sum\limits_{i=1}^3z_{k,i}(\vec{q})p_i+u_k(\vec{q})\Bigr) \prod\limits_{j\neq k}\frac{\alpha-q_j}{q_k-q_j},\quad \lbrace h(\alpha),h(\beta)\rbrace=0
\end{equation}
where
\begin{equation}\label{symm1}
S(x)=a^2(x),\quad z_{i,j}(\vec{q})=-\delta\frac{a(q_i)a(q_j)}{q_i-q_j},\;i\neq j,\; z_{i,i}(\vec{q})=0,\quad a(x)=a_3x^3+a_2x^2+a_1x+a_0,
\end{equation}
$$
u_1(\vec{q})=u(q_1,q_2,q_3),\;u_2(\vec{q})=u(q_2,q_3,q_1),\;u_3(\vec{q})=u(q_3,q_1,q_2)
$$
and
$$
u(x,y,z)=-\frac{\delta^2}{4}\Bigl(3S(x)(\frac{1}{(x-y)^2}+\frac{1}{(x-z)^2})-S'(x)(\frac{1}{x-y}+\frac{1}{x-z})+\frac{1}{5}S''(x)\Bigr)
$$
\end{statement}

\paragraph{Remark.}
1) In fact, $u(x,y,z)$ is defined up to a quadratic polynomial $b(x)=b_0x^2+b_1x+b_2$, but one can exclude it by the shifts $h_i\rightarrow h_i-b_i.$

2) In the case $a_3=0,$ that is if $a(x)$ is a quadratic polynomial,
the canonical transformation
$$p_i\rightarrow p_i+\frac{\partial}{\partial_{q_i}}G(q_1,q_2,q_3),\;G(x,y,z)=\frac{\delta}{2}\log((x-z)(x-y)(y-z))$$
allows to exclude the potential $u$ completely.
In this case, the family (\ref{ha2}) takes the form
\begin{equation}\label{symm_g}
h(\alpha)=\sum\limits_{k=1}^3 \Bigl( a^2(q_k)p_k^2+ a(q_k)\delta\sum\limits_{i\neq k}\frac{a(q_k)p_k-a(q_i)p_i}{q_k-q_i}\Bigr)\prod\limits_{j\neq k}\frac{\alpha-q_j}{q_k-q_j}.
\end{equation}
\medskip
%


Using relation $h(\alpha)=\sum\limits_{k=0}^{2}h_k\alpha^{2-k}$ we obtain Hamiltonians $h_0,h_1,h_2$ in involution: $\lbrace h_0,h_1 \rbrace=0, \lbrace h_1,h_2 \rbrace=0, \lbrace h_2,h_0 \rbrace=0,$
therefore the system (\ref{symm_g}) is a Liouville integrable. Let us discuss some applications of this system to the problems of hydrodynamic type.
To this end, we introduce three times $t,\tau,\xi$ which describe the full evolution of the system:
\begin{equation}\label{times}
A_t=\lbrace A,h_0 \rbrace,\quad A_{\tau}=\lbrace A,h_1 \rbrace,\quad A_{\xi}=\lbrace A,h_2 \rbrace.
\end{equation}
and denote $I_1=q_1+q_2+q_3,\; I_2=q_1q_2+q_2q_3+q_3q_1,\; I_3=q_1q_2q_3.$

One can easily prove by computing $\frac{\partial h_1}{\partial p_k}+(I_1-q_k)\frac{\partial h_0}{\partial p_k}$ that any solution of (\ref{times}) satisfies the hydrodynamic type system
\begin{equation}\label{GTa}
q_{k,\tau}+(I_1-q_k)q_{k,t}=\frac{\delta}{2}a''(q_k)a(q_k)-a^2(q_k)\delta\prod\limits_{j\neq k}\frac{1}{q_k-q_j},\quad I_{1,\tau}=-I_{2,t}
\end{equation}
Quite similarly, the computation of $\frac{\partial h_2}{\partial p_k}-\frac{I_3}{q_k}\frac{\partial h_0}{\partial p_k}$  brings to another hydrodynamic type system
\begin{equation}\label{GTb}
q_{k,\xi}-\frac{I_3}{q_k}q_{k,t}=-a(q_k)\delta\sum\limits_{(i,j,k)=(1,2,3)} \frac{a(q_i)q_j}{(q_i-q_k)(q_i-q_j)},\;  \quad I_{1,\xi}=I_{3,t},
\end{equation}

A system (\ref{GTa}) is a generalization of the celebrated Gibbons--Tsarev system \cite{GT}. Some systems analogous to (\ref{GTa},\ref{GTb}) were considered in \cite{ferap} and recently in \cite{sok}.
Both systems can be generalized to the case of any dimension.

We do not know at the moment the separation of variables for system (\ref{symm_g}) and whether it is solvable.

But we will show that in the special symmetric case $a(x)=1$ the system is solvable. In the rest of the section,
we consider this case in detail and obtain its general solution. To this end, we write the system (\ref{q-stack}) with $a(x)=1$ in a `separated'
form (in $\delta=0$ case this system is separated indeed and coincides with the St\"ackel one) as follows
\begin{equation}\label{symm}
q_k^2h_0+q_kh_1+h_2-p_k^2-\delta \sum\limits_{j\neq k}\frac{p_k-p_j}{q_k-q_j}=0,\; k,j=1,2,3.
\end{equation}

We solve system (\ref{symm}) linear with respect to $h_k$, denote $h_k=H_k+V_k\delta,\;k=0,1,2$ and obtain
\begin{equation}\label{symmh2}
H_0=\sum\limits_k p_k^2\prod\limits_{j\neq k}\frac{1}{q_k-q_j},\;H_1=\sum\limits_k p_k^2(q_k-I_1)\prod\limits_{j\neq k}\frac{1}{q_k-q_j},\;H_2=\sum\limits_k p_k^2\prod\limits_{j\neq k}\frac{q_j}{q_k-q_j},
\end{equation}
\begin{equation}\label{symmV}
V_0=0,\; V_1=\sum\limits_k p_k\prod\limits_{j\neq k}\frac{1}{q_k-q_j},\; V_2=\sum\limits_k p_k (2q_k-I_1)\prod\limits_{j\neq k}\frac{1}{q_k-q_j}.
\end{equation}
We introduce two more functions
\begin{equation}\label{symmVa}
V_3=\sum\limits_k p_k (I_1^2-2I_2-2q_k^2)\prod\limits_{j\neq k}\frac{1}{q_k-q_j},\;V_4=\sum\limits_k p_k
\end{equation}
to obtain a Poisson algebra with 7 generators
$H_0,H_1,H_2,V_1,V_2,V_3,V_4:$
$$
\lbrace H_0,H_1\rbrace=0,\quad \lbrace H_1,H_2\rbrace =0,\quad\lbrace H_2,H_0\rbrace =0,
$$
$$
\lbrace H_0,V_1\rbrace=0,\quad \lbrace H_0,V_2\rbrace=0,\quad\lbrace H_0,V_3\rbrace=0,\quad\lbrace H_0,V_4\rbrace=0,\quad
$$
$$
\lbrace V_1,V_2\rbrace=0,\;\lbrace V_1,V_3\rbrace=0,\;\lbrace V_1,V_4\rbrace=0,\;\lbrace V_2,V_3\rbrace=2V_1,\; \lbrace V_2,V_4\rbrace=V_1,\;\lbrace V_3,V_4\rbrace=2V_2,\;
$$
$$
\lbrace H_1,V_1\rbrace=-V_1^2,\; \lbrace H_1,V_2\rbrace=-V_1V_2,\; \lbrace H_1,V_3\rbrace=-V_2^2,\; \lbrace H_1,V_4\rbrace=2H_0,\;
$$
$$
\lbrace H_2,V_1\rbrace=\lbrace H_1,V_2\rbrace,\;\lbrace H_2,V_2\rbrace=-V_1V_3,\; \lbrace H_2,V_3\rbrace=-V_2(V_3+2V_4)-2H_1,\;\lbrace H_2,V_4\rbrace=H_1.
$$

The main reason that the system is solvable is not only Liouville integrability but the fact that $\lbrace h_0,V_i \rbrace =0,\; i=1,\dots ,4.$
This allows to introduce the notation $V_i=v_i(\tau,\xi),$ and to find the momenta $p_1,p_2,p_3$ from the linear system $V_1=v_1(\tau,\xi),\; V_2=v_2(\tau,\xi),\; V_4=v_4(\tau,\xi).$
The variables $q_i$ are found then by use of only first Hamiltonian equation
$$q_{i,t_{\alpha}}=\frac{\partial h_{\alpha}}{\partial p_i}$$
after substituting the obtained momenta.
In order to determine $q_i$ it is convenient to consider the dynamical systems
\begin{equation}\label{int_qst}
\frac{\partial I_j}{\partial t_{\beta}}=\lbrace I_j,H_{\beta} \rbrace,\quad j=1,2,3,\; \beta=0,1,2,\quad t_{\beta=0}=t,\; t_{\beta=1}=\tau,\; t_{\beta=2}=\xi.
\end{equation}

First, we find the $t$-dynamics of $I_k:$
\begin{equation}\label{t-dyn}
I_1=-2v_1(\tau,\xi)t+c_1(\tau,\xi),\;I_2=-v_1(\tau,\xi)(-v_1(\tau,\xi)t^2+c_1(\tau,\xi)t)+v_2(\tau,\xi)t+c_2(\tau,\xi),
\end{equation}
$$
I_3=-\frac{1}{18}c_1(\tau,\xi)^3-v_1(\tau,\xi)v_2(\tau,\xi)t^2+v_2(\tau,\xi)c_1(\tau,\xi)t+v_3(\tau,\xi)t+c_3(\tau,\xi).
$$

Then step by step we restore  the $\tau$- and $\xi$-dependence using the additional identities:
$$
V_3=v_3(\tau,\xi),\; h_i=\tilde{h}_i
$$
where $\tilde{h}_i$ are constant values of Hamiltonians $h_i.$

Finally, we obtain the general solution of the quasi-St\"ackel system under consideration as the zeroes of the cubic equation
$$
 (q-q_1)(q-q_2)(q-q_3)=q^3-I_1q^2+I_2q-I_3=0
$$
with the coefficients
$$
I_1=\frac{-2(t-t_0)+\lambda r}{\tau_0-\tau}-\lambda\tau_1,\quad
I_2=\frac{(t-t_0-\frac{1}{2}\lambda r)^2}{(\tau_0-\tau)^2}-\frac{1}{4}\lambda^2(\tau_0-\tau)^2+\lambda^2 \tau_2(\tau_0-\tau)-\lambda^2\tau_1^2,
$$
$$
I_3=\lambda\tau_1(\frac{t-t_0-\frac{1}{2}\lambda r}{\tau_0-\tau}+\tau_1\lambda)^2-\lambda^2(t-t_0)\tau_2+\frac{1}{2}\lambda^3r \tau_2 -(\lambda^3\tau_1\tau_2+\delta)(\tau_0-\tau)-\frac{1}{4}\lambda^3(\tau_1-2\tau_3)(\tau_0-\tau)^2
$$
where notations are used
$$
f(\lambda\xi)=C_1+C_2\lambda\xi+C_3e^{-\lambda\xi}+C_4e^{\lambda\xi},\quad \tau_0=f(\lambda\xi),\;\tau_1=f'(\lambda\xi),\;\tau_2=f''(\lambda\xi),\;\tau_3=f'''(\lambda\xi),\;
$$
$$
r=\lambda\xi ((\tau_1-\tau_3)^2+\tau_3^2-\tau_2^2)+\tau_0(\tau_1-\tau_3)+\tau_2(3\tau_1-2\tau_3),
$$
$$
h_0=\frac{1}{4}\lambda^2,\; h_1=-\frac{1}{2}C_2\lambda^3 ,\;h_2=\lambda^4 (C_3C_4+\frac{1}{4}C_2^2).
$$
This solution depends on the full set of integration constants $t_0,C_1,C_2,C_3,C_4,\lambda$.

\subsection{Non-symmetric case}
In this case we obtain one-dimensional St\"ackel system (recall that we can assume that $H_i$ are constant)
$$
x^2H_0+xH_1+H_2-S(x)p_1^2=0.$$
The action in this case is of the form
${\bf S}=\int\, dx\, \sqrt{\frac{x^2H_0+xH_1+H_2}{S(x)}}$.

The remaining two-dimensional (quasi-St\"ackel) system reads
$$
y^2H_0+yH_1+H_2-S(y)p_2^2-\delta\frac{\sqrt{S(y)}\sqrt{S(z)}}{y-z}p_3+\frac{\delta^2}{4}\Bigl(3\frac{S(y)}{(y-z)^2}-\frac{S'(y)}{y-z}+\frac{1}{10}S''(y)\Bigr)=0,
$$
$$
z^2H_0+zH_1+H_2-S(z)p_3^2-\delta\frac{\sqrt{S(y)}\sqrt{S(z)}}{z-y}p_2+\frac{\delta^2}{4}\Bigl(3\frac{S(z)}{(z-y)^2}-\frac{S'(z)}{z-y}+\frac{1}{10}S''(z)\Bigl)=0.
$$
where $S$ is a polynomial of sixth degree \cite{MarSok1}.

Notice that Hamiltonians $H_i$ defined by these systems are three-dimensional, that is, they depend on all variables $q_i,p_i$, $i=1,\dots ,3.$
At the moment, only the partial separation of variables is known for this system \cite{MarSok2}.

In this article we considered a classification of three-dimensional quasi-St\"ackel systems.
We obtain only two families: - symmetric three-dimensional system and non-symmetric ones. In the case $a(x)=1$ the symmetric system is solvable and we obtain the general solution of this system.
The calculations become very involved for the case $n>3$ and at the moment one can hardly expect to obtain a full classification even in the $n=4$ case.
On the other hand, system (\ref{symm}) admits a straightforward generalization for the case $n>3.$
Preliminary result is that in $n=4$ case we obtain $\lbrace h_i,h_j \rbrace=0,\; i,j=0,1,2,3$ except for the relation
$\lbrace h_2,h_3\rbrace\neq 0.$ This is a surprising fact and we have no Liouville integrability in this case.

{\bf Acknowledgments.} The author thanks V. E. Adler, V. V. Sokolov, and A. B. Shabat for the useful
discussions.
This research was supported by the Program for Supporting Leading Scientific Schools (Grant No. NSh 3139.2014.2) and the Russian Foundation for Basic Research (Grant No. 13-01-00402).

\end{document}